\documentclass[12pt]{article}
\input psfig.sty
\newcommand{\be}{\begin{equation}}
\newcommand{\ee}{\end{equation}}
\newcommand{\bea}{\begin{eqnarray}}
\newcommand{\eea}{\end{eqnarray}}
\newcommand{\sptwo}{1.4}
\newcommand{\doublespace}{\edef\baselinestretch{\sptwo}\Large\normalsize}
\newcommand{\newsection}[1]{\section{#1}\setcounter{equation}{0}}

\newcounter{newapp}
\begin{document}
\begin{titlepage}
\vspace*{1.0in}
\begin{center}
{\large\bf The Effective K\"ahler Potential\\
 And Extra Space-Time Dimensions}
\end{center}
\vspace{0.25in}
~\\
\begin{center}
{\bf T.E. Clark and S.T. Love}\\
{\it Department of Physics\\ 
Purdue University\\
West Lafayette, IN 47907-1396}
~\\
~\\
\end{center}
\vspace{1in}
\begin{center}
{\bf Abstract}
\end{center}

The effects of extra space-time dimensions on the Wilsonian effective K\"ahler potential and the perturbative one loop effective K\"ahler potential are determined within the framework of an Abelian gauge theory with $N=2$ supersymmetric field content.  The relation between the K\"ahler metric and the effective gauge couplings which leads to the absence of radiative corrections to the K\"ahler potential is expressed as a function of the radius of compactification of a fifth dimension.  In general, the quantum corrections to the low energy K\"ahler potential are shown to grow with this radius reflecting the underlying higher dimensional nature of the theory.
\end{titlepage}
\pagebreak

\doublespace

\newsection{Introduction}

As a consequence of the large degree of symmetry present in supersymmetric theories, the form of their low energy Wilsonian effective actions is highly constrained. For instance,  when $N=1$ SUSY is combined with the required holomorphic dependence on the fields and parameters of a model, the superpotential can be completely (non-perturbatively) determined even in the presence of strongly interacting gauge fields \cite{S}. Furthermore, for extended $N=2$ supersymmetric theories, the entire low energy action can be constructed in terms of a single holomorphic prepotential.  As in the $N=1$ superpotential case, this prepotential is exactly determined using the symmetries, holomorphicity and duality properties of the model in question \cite{SW}.  The simplest $N=2$ gauge theory consists of an $N=1$ $U(1)$ gauge superfield $V$ and its neutral chiral $\varphi$ and antichiral $\bar\varphi$ superfield $N=2$ partners.  In terms of these $N=1$ superfields, the low energy effective action has the form
\bea
\Gamma_{N=2} [\varphi, \bar\varphi, V] &=& \frac{1}{8\pi i}\left(\int dV \left[ {\cal F}_\varphi(\varphi)\bar\varphi-
\bar{\cal F}_{\bar\varphi}(\bar\varphi)\varphi \right]\right. \cr
 & & \left. \right. \cr
 & & \left.+\frac{1}{2} \int dS
{\cal F}_{\varphi\varphi}(\varphi)W^{\alpha}W_\alpha -\frac{1}{2} \int d\bar S \bar{\cal F}_{\bar\varphi\bar\varphi}(\bar{\varphi})
\bar W_{\dot\alpha}\bar W^{\dot\alpha} \right) .\cr
 & & 
\label{N=2action}
\eea
Here subscripts denote differentiation with respect to that variable so that, for example, ${\cal F}_\varphi = \frac{\partial {\cal F}}{\partial \varphi}$. The chiral field strength is defined by $W_\alpha =-\frac{1}{4}\bar D\bar D e^{-2V}D_\alpha e^{2V}= -\frac{1}{2}\bar D\bar D D_\alpha V$ with a similar definition for the anti-chiral field strength: $\bar W_{\dot\alpha}= -\frac{1}{2}DD\bar D_{\dot\alpha}V$.  The holomorphic prepotential ${\cal F}(\varphi)$ then determines the K\"ahler potential $K(\varphi,\bar{\varphi})={\rm Im}({\cal F}_\varphi (\varphi) \bar\varphi)$ and hence the K\"ahler metric $g(\varphi,\bar{\varphi})=K_{\varphi\bar{\varphi}}={\rm Im}{\cal F}_{\varphi\varphi}$ as well as the effective gauge couplings: ${\cal F}_{\varphi\varphi}~,~\bar{\cal F}_{\bar\varphi\bar\varphi}$. The self-radiative corrections to this general $N=2$ SUSY Abelian gauge theory action were found to vanish \cite{S}\cite{CL1}\cite{SV}.

Extra compact space dimensions are a common attribute of many currently proposed fundamental theories.  Concomitant with these is the tower of Kaluza-Klein (K-K) field excitations whose masses are integer multiples of the inverse compactification radius $R$.  In models where $R$ is the order of the inverse Planck scale or smaller, the K-K modes unconsequentially decouple from the low energy physics.  On the other hand, for radii much larger than the inverse Planck scale, low energy physics can be considerably affected by the propagation of fields into the large radii extra dimensions.  At the very least, theories formulated in higher dimensions tend to have a more divergent short distance behavior which leads to the breakdown of renormalizability. Thus an ultraviolet cutoff $\Lambda$ is necessary for a consistent interpretation of such field theories.  For large compactification radii, $\Lambda R >1$, the K-K modes with mass below the cutoff can effect the low energy physics \cite{A}-\cite{AQ}. The resulting theoretical and phenomenological consequences include, for example, the possibility of introducing \cite{SSCSS}-\cite{DGD} a consistent and calculable mechanism of supersymmetry breaking and/or communicating the supersymmetry breaking between different four dimensional boundaries \cite{Ho}\cite{MP} and could also be instrumental in the description of the strong coupling limit of string theory \cite{HoW}.  In addition, their influence on the running of the gauge coupling and its effect on the scale of unification has been investigated \cite{DDG}.  It has even been suggested that they could lead to possible experimentally accessible modifications to gravitational interactions in the millimeter range \cite{ADD}.

In four dimensions, the general low energy action containing terms through two space-time derivatives for a $N=1$ supersymmetric theory with the $N=2$ field content of an Abelian gauge superfield $V$ and a neutral (anti-)chiral field $(\bar\varphi)~\varphi$, has the form
\bea
\Gamma [\varphi, \bar\varphi, V] &=& \int dV K(\varphi, \bar\varphi) +\frac{1}{2} \int dS
{f}(\varphi)W^{\alpha}W_\alpha +\frac{1}{2} \int d\bar{S} \bar{f}(\bar\varphi)
\bar{W}_{\dot\alpha}\bar{W}^{\dot\alpha}  \cr
 & & \cr
 & & +\int dS P(\varphi) +\int d\bar S \bar P(\bar\varphi)
+\kappa \int dV V .
\label{zeroaction}
\eea
Here the K\"ahler potential, $K(\varphi,\bar\varphi)$, the effective gauge couplings, $f(\varphi)$ and $\bar f(\bar\varphi)$, and the superpotentials, $P(\varphi)$ and $\bar P(\bar\varphi)$, are arbitrary, unrelated functions of their respective arguments. The linear in $V$ Fayet-Iliopolous term has an arbitrary coupling constant, $\kappa$, which as a consequence of gauge invariance is necessarily field independent. Independent radiative corrections vanish for the superpotential in accord with the non-renormalization theorem in this topologically trivial model.  Likewise, the Fayet-Iliopolous term has no independent additive radiative corrections.  Hence each of these terms remain zero if they initially vanish.  Finally, while the effective gauge couplings $f(\varphi)$ and $\bar f(\bar\varphi)$ also remain uncorrected, the K\"ahler potential does receive radiative corrections.  Setting the Fayet-Iliopolous terms to zero, these quantum effects are given by the Wilson renormalization group equations (WRGE) \cite{W}\cite{HW}\cite{CHL} for $K$, $f$($\bar f$) and $P$($\bar P$) as \cite{CL1}
\bea
{\partial K\over \partial t} &=& 2K-(1-\gamma)\varphi K_\varphi -(1-\gamma)\bar\varphi K_{\bar\varphi}+{1\over 8\pi^2}\ln{\left[{g\over f+\bar f}\right]}\cr
 & & \cr
{\partial f\over \partial t} &=& 2\gamma_V  f -(1-\gamma)\varphi f_\varphi \quad ; \quad
{\partial \bar f\over \partial t} = 2\gamma_V  \bar f -(1-\gamma)\bar\varphi \bar f_{\bar\varphi}\cr
 & & \cr
{\partial P\over \partial t} &=& 3P-(1-\gamma ) \varphi P_\varphi \quad ; \quad
{\partial \bar P\over \partial t} = 3\bar P-(1-\gamma ) \bar\varphi \bar P_{\bar\varphi} ,
\label{wrge}
\eea
where all quantities have been made dimensionless by scaling with the
loop momentum cutoff $\Lambda$, so that, for example $K\rightarrow K /\Lambda^2$ and $\varphi\rightarrow \varphi /\Lambda$. The result was obtained using the $R_\xi$ gauge fixing action term
\be
\Gamma_{\xi} [V] = \xi \int dV  DDV\bar D\bar D V
\ee
and included all graphical contributions through two space-time derivatives and arbitrary powers of the fields. In this simple Abelian model, the gauge fixing parameter $\xi$ receives no radiative corrections nor contributes to the other functions.

Since the fields are rescaled according to their wavefunction renormalization in order to maintain a canonical kinetic energy term as the short distance degrees of freedom are integrated out of the theory, the chiral field anomalous dimension $\gamma $ can be found by evaluating the renormalization group equation for the K\"ahler metric, $g=K_{\varphi\bar\varphi}$,  at $\varphi=0=\bar\varphi$ where it is normalized as $g\vert_{\varphi =0=\bar\varphi}={1}$.  Likewise, the photon anomalous dimension, $\gamma_V$, is secured using the renormalization group equation for $f$ and $\bar f$ evaluated at zero fields $\varphi =0=\bar\varphi$ along with the normalizations $f\vert_{\varphi =0}={1\over 2}=\bar f\vert_{\bar\varphi =0}$ . So doing, one finds that $\gamma_V =0$ and 
\be
2\gamma = -{1\over 8\pi^2}\left[ g_{\varphi\bar\varphi} -g_\varphi g_{\bar\varphi} +f_\varphi \bar f_{\bar\varphi} \right]\bigg\vert_{\varphi =0=\bar\varphi} .
\ee
If the initial choice of K\"ahler potential and effective gauge coupling satisfies the $N=2$ supersymmetric relation $g=f+\bar f$, then the anomalous dimension of the chiral field vanishes and the Wilson renormalization group equation for $K$ reduces to that of naive scaling. There are no radiative corrections to the Wilson effective action. 

One can also compute the corrections using ordinary perturbation theory. The one loop correction to the dimensionless effective K\"ahler potential is found to be simply given by
\be
\delta K_{\rm 1-loop}= -\frac{1}{16\pi^2}\ln{\left[\frac{g}{f+\bar f}\right]}.
\ee
Just as was the case in the WRGE analysis, the $N=2$ supersymmetric relation $g=f+\bar f$ leads to no one-loop radiative corrections \cite{S}, \cite{GRvU}-\cite{PW}.

The purpose of this paper is to determine the effects of an extra compact dimension on the supersymmetric Wilsonian effective action as well as its effects on the perturbative one-loop effective action.  Starting with an $N=1$ SUSY theory with a compact fifth dimension and Fourier expanding the fields over the circular fifth dimension results in the four dimensional theory with the $N=2$ SUSY field content of the above described zero mode fields, $V,~\varphi$, and  $\bar\varphi$, as well as the infinite tower of massive K-K $N=2$ SUSY multiplets, $\{ V_n,~\phi_n,~\bar\phi_n \}$, for $n$ any non-zero (positive and negative) integer.  The resultant four dimensional action can thus be written
as an expansion in powers of the K-K tower fields as,
\be
\Gamma = \sum_{l=0}^\infty \Gamma_l .
\ee
Here $\Gamma_0$ contains zero mode fields only. Due to momentum conservation in the direction of the fifth dimension, $\Gamma_1 =0$, while $\Gamma_2$ contains 2 K-K tower fields, and so on.  We seek to construct the Wilson renormalization group equation for the zero mode field action, $\Gamma_0 [V, \varphi, \bar\varphi ]$ after inclusion of the radiative effects of K-K towers.  In addition, the perturbative one loop radiative corrections to $\Gamma_0$ will be calculated.  

In both cases, the quantum corrections arise from the self-radiative contributions from $\Gamma_0$ as well as from radiative corrections of $\Gamma_2$ to $\Gamma_0$.   In either case, the form of $\Gamma_2$ is required for the determination of the quantum corrections.  In a complete Wilson renormalization group analysis, the evolution of $\Gamma_2$ is determined by itself and $\Gamma_4$ and in turn $\Gamma_4$ is determined by contributions from $\Gamma_6$ and so on.  Similarly, higher order perturbative contributions to $\Gamma_0$ come from higher numbers of K-K tower fields in the action. Since such an infinitely iterative procedure is beyond our calculational means, we must make additional assumptions regarding the form of the $\Gamma_0$ and $\Gamma_2$ actions before we can proceed. First of all, we neglect all higher derivative terms than those appearing in the kinetic energy. This constitutes the next to leading order in a derivative expansion of the action and allows for the determination of the anomalous dimensions. It thus constitutes an improvement beyond the local potential approximation \cite{HH} which is possible \cite{CLFP} because the kinetic terms arise from a K\"ahler potential. Thus for $\Gamma_0$, we use Eq.(\ref{zeroaction}) without the Fayet-Iliopoulos term ($\kappa = 0$). A similar form is assumed for the $\Gamma_2$ action except that explicit compactified radius dependent mass terms are included for the $N=2$ K-K tower fields. Thus we take  
\bea
\Gamma_2 &=& \sum_{n=-\infty \atop n\neq 0}^{+\infty} \left\{ \int dV \left[g (\varphi,\bar\varphi ) \bar\phi_n \phi_n +M^2_n V_n V_{-n}\right]\right.\cr
 & & \left. \right. \cr
 & & \left.\int dS \left[\frac{1}{4} f(\varphi) W^\alpha_n  W_{-n\alpha} +m_n(\varphi) 
\phi_n \phi_{-n}\right]\right. \cr
 & & \left. \right. \cr
 & &\left.+ \int d\bar{S}\left[\frac{1}{4} \bar{f} (\bar\varphi) \bar{W}_{n\dot\alpha} \bar{W}^{\dot\alpha}_{-n} + \bar{m}_n(\bar\varphi)\bar\phi_n \bar\phi_{-n}\right]\right\}
\label{KKaction}
\eea 
The K-K vector field mass term is given by $M_n^2 =\frac{n^2}{R^2}$ while the K-K (anti-) chiral field mass term is ($\bar{m}_n(\bar\varphi)=\frac{n}{R}-i\bar{P}_{\bar{\varphi}\bar{\varphi}}$) $m_n(\varphi)=\frac{n}{R}+iP_{\varphi\varphi}$, where $n=\pm 1, \pm2, \ldots, \pm\infty$. 
The form of the action $\Gamma_2$ corresponds to a radical truncation of all generally possible terms.  It has the property that all modes have the same, isotropic four dimensional action coefficients.  Even in the large $R$ limit, it is only the terms with momentum in the fifth dimension that are treated anisotropically.  Of course, this is by no means a justification of the truncation which can only be achieved by a detailed analysis of the compactification of the more fundamental SUSY theory involved.  This is quite a lengthy task and is beyond the scope of this present investigation.

\newsection{The Wilson Effective K\"ahler Potential}

The Wilson renormalization group equation governing the evolution of the zero mode K\"ahler potential is determined by integrating out the degrees of freedom with momentum in an infinitesimal four dimensional shell about $\Lambda (t) =e^{-t}\Lambda$.  Although we are summing over all K-K modes in this shell, those with mass above the scale $\Lambda (t)$ decouple. As the sum over these modes takes place, the action is rescaled according to the wavefunction renormalization of the fields. Fixing the zero field metric at the scale $\Lambda (t)$ to its canonically normalized value of one, $g\vert_{\varphi =0=\bar\varphi}=1$,  and further rescaling all dimensionful quantities by the appropriate factors of $\Lambda$ needed to render them dimensionless, the Wilson renormalization group equation for the zero mode K\"ahler potential is obtained as
\bea
\frac{\partial K}{\partial t} &=& 2K -\left( 1-\gamma \right) \varphi K_\varphi -\left( 1-\gamma \right) \bar\varphi K_{\bar\varphi} \cr
& &+\frac{1}{16\pi^2}\sum_{n=-\infty}^{+\infty} \ln{\left[\left( \frac{g^2 + r+\frac{n^2}{R^2}}{[(f+\bar{f})+ \frac{n^2}{ R^2}]^2}\right)\left(\frac{[1+ \frac{n^2}
{R^2}]^2}{1 + r(0)+\frac{n^2}{R^2}}\right)\right]}.
\eea
Here the dimensionless mass coefficient is  $r=P_{\varphi\varphi}\bar{P}_{\bar\varphi\bar\varphi}$, with $r(0)$ the value of the dimensionless mass parameter at zero field, $r(0)=r|_{\varphi =0= \bar\varphi }$. The effective gauge couplings and superpotential remain uncorrected and their Wilson renormalization group equations are once again given in equation (\ref{wrge}). The sum over the K-K modes may now be performed yielding
\bea
\frac{\partial K}{\partial t} &=& 2K -\left( 1-\gamma \right) \varphi K_\varphi -\left( 1-\gamma \right) \bar\varphi K_{\bar\varphi} \cr
 & &\quad +\frac{1}{8\pi^2}\ln{\left[ \frac{\sinh{\left[\pi  R \sqrt{g^2 + r}\right]}}{\sinh{\left[\pi  R \sqrt{1 + r(0)}\right]}}\frac{\sinh^2{[\pi R ]}}{\sinh^2{\left[\pi R \sqrt{f +\bar{f}}\right]}}\right]},
\label{wrge2}
\eea
where $g$ and $f$, $\bar{f}$ have been normalized at zero field to be 1 and 1/2, 1/2,  respectively.

The four dimensional limit, $ R\rightarrow 0$, reproduces the Wilson renormalization group equation (\ref{wrge}) obtained previously.  For $ R >> 1$, the Wilson renormalization group equation for $K$ reduces to 
\bea
\frac{\partial K}{\partial t} &=& 2K -\left( 1-\gamma \right) \varphi K_\varphi -\left( 1-\gamma \right) \bar\varphi K_{\bar\varphi}\cr
 & &\quad +\frac{2\pi  R}{16\pi^2} \left[\sqrt{g^2 + r} -2\sqrt{f + \bar{f}}-\sqrt{1+r(0)} +2 \right].
\eea
This can be recast by further rescaling the fields and the K\"ahler potential according to their five dimensional engineering dimensions, $\varphi \rightarrow \frac{\varphi}{\sqrt{2\pi R}}~,~ 
\bar\varphi \rightarrow\frac{\bar\varphi}{\sqrt{2\pi R}}~,~
K \rightarrow \frac{1}{2\pi R}K$,
as
\bea
\frac{\partial K}{\partial t} &=& 3K -\left( \frac{3}{2}-\gamma \right) \varphi K_{\varphi} -\left( \frac{3}{2}-\gamma \right) \bar\varphi K_{\bar\varphi} \cr
 & &\qquad +\frac{1}{16\pi^2} \left[\sqrt{g^2 + r} -2\sqrt{f + \bar{f}}-\sqrt{1+r(0)}+2 \right].
\eea
This result can also be obtained directly in five dimensions by integrating over the fifth dimension momentum variable rather than performing a Fourier series. 

For $ R\neq 0$, the radiative corrections to the K\"ahler potential do not, in general, vanish at the four dimensional, $N=2$ relation between the metric and effective gauge coupling. Rather one finds that, with $r =0$, the logarithm on the right hand side of equation (\ref{wrge2}) vanishes resulting in no radiative correction to the K\"ahler potential, $K$, provided 
\be
\frac{\sinh{\left[\pi  R\right]}\sinh{\left[\pi  R g\right]}}{\sinh^2{\left[\pi  R \sqrt{f+\bar{f}}\right]}}= 1
\label{norad}
\ee
When this condition is satisfied, the Wilson renormalization group equation yields the naive scaling equation for the K\"ahler potential. 

The four dimensional $N=2$ relation between metric and effective gauge couplings that leads to the absence of radiative corrections is $g=(f+\bar{f})$, which is also the  $ R \rightarrow 0$ limit of equation (\ref{norad}). In figure 1, we plot the relation between $g$ and $f+\bar{f}$ for different compactified radii, $R$, values which leads to no radiative corrections. We see that, due to the exponential dependence on $ R$, the no radiative correction condition quickly deviates from the four dimensional relation and rapidly approaches the five dimensional relation. For $f+\bar{f}>1$, the values for $g$ lie on top of the five dimensional values within the resolution of the figure for $R=1, ~2, ~10$.  For $f + \bar{f}<1$, the curves for $R=0,~ 0.1,~ 0.2$ essentially coalesce while those for $R=1,~2,~ 10$ have differing behavior from each other and from the four dimensional result.  In general, for fixed $f+\bar{f}$, the radiative corrections vanish for smaller $g$ values than that of the four dimensional case and even take on exponentially small values, $g \approx 2\pi R e^{-\pi R}(f+\bar{f})$, in the strong gauge coupling limit when $f+\bar{f} << 1$.
\begin{figure}
\psfig{file=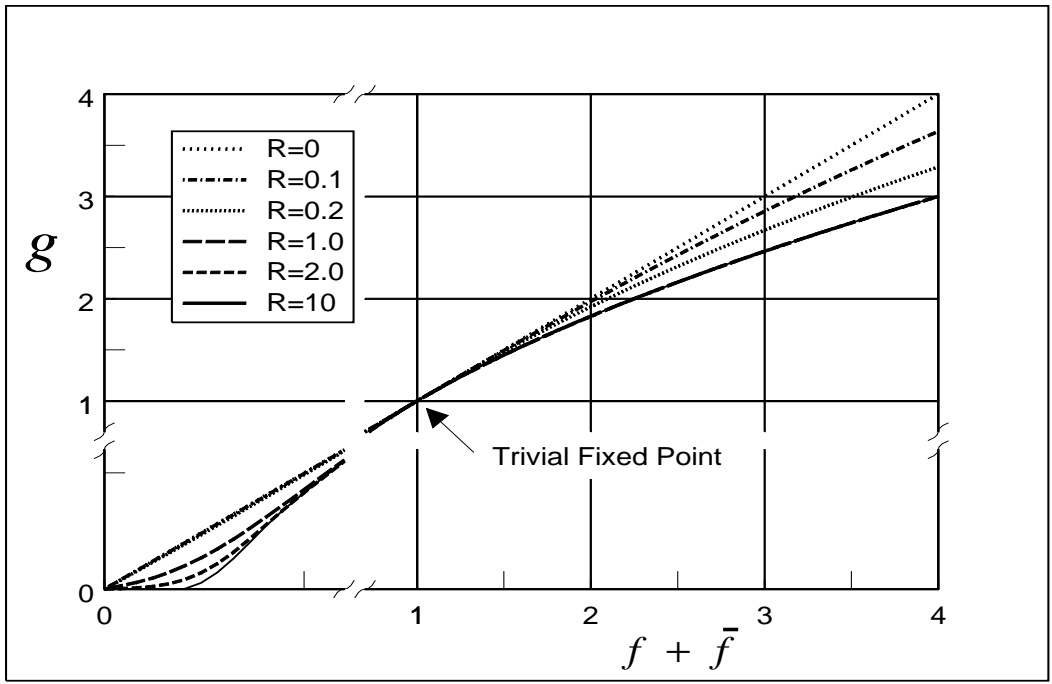,width=6.0in,height=4in}
\caption{The relation between $g$ and $f+\bar{f}$ resulting in the absence of radiative corrections for various compactification radii.}
\end{figure}

Being a trivial theory in four and above dimensions, the zero radiative correction line is an attractive relation between $g$ and $f+\bar{f}$ in the sense that the renormalization group flow in the $(g,f,\bar{f})$ space tends towards the no radiative correction line, reaching it in the infrared at the single trivial point $g=1$ and $f=\frac{1}{2}=\bar{f}$.  This can be seen by considering the evolution of the ratio
\be 
\rho \equiv 1+\xi \equiv \frac{\sinh{[\pi  R ]}\sinh{[\pi R g ]}}{\sinh^2{\left[\pi R \sqrt{f+\bar{f}}\right]}},
\ee
where $\xi\vert_{\varphi=0=\bar\varphi} =0$.  Applying the Wilson renormalization group equations (\ref{wrge2}) and (\ref{wrge}) gives
\bea
\frac{\partial \rho}{\partial t} &=& \left(2\gamma g \pi R \coth{[\pi R g ]}\right) \rho -\left( 1-\gamma \right) \varphi \rho_{\varphi} -\left( 1-\gamma \right) \bar\varphi \rho_{\bar\varphi} \cr
 & &+\frac{\pi R}{8\pi^2}\coth{[\pi R g ]}\left[ \rho_{\varphi\bar\varphi} -\frac{\rho_\varphi \rho_{\bar\varphi}}{\rho} \right] .
\eea
Note that the anomalous dimension of the chiral fields, $\gamma$, can be expressed in terms of $\xi$ as
\be
2\gamma = -\frac{1}{8\pi^2}\left[ \rho_{\varphi\bar\varphi} -\rho_{\varphi} \rho_{\bar\varphi} \right]\vert_{\varphi =0=\bar\varphi} 
= -\frac{1}{8\pi^2}\left[ \xi_{\varphi\bar\varphi} -\xi_{\varphi} \xi_{\bar\varphi} \right]\vert_{\varphi =0=\bar\varphi} .
\ee
For small $\xi$, $g$ is slightly off the fixed relation no radiative correction curve.  Hence, writing $g=g_0 +\Delta g$, where $g_0$ satisfies the no radiative correction condition, and expanding equation (\ref{norad}) yields
\be
\Delta g = \frac{\tanh{[\pi R g_0 ]}}{\pi R} \xi .
\ee
Using Eq. (\ref{norad}), the evolution of $\xi$ is given by
\be
\frac{\partial \xi}{\partial t} = -\varphi \xi_\varphi - \bar\varphi \xi_{\bar\varphi} +\frac{\pi R}{8\pi^2}\coth{\left[\pi R g_0\right]}\left[\xi_{\varphi\bar\varphi}
-g_0 \xi_{\varphi\bar\varphi}\vert_{\varphi =0= \bar\varphi}\right] +O(\xi^2).
\ee
As the theory flows into the infrared, $t>0$, the values of the scaled fields, $\varphi\rightarrow e^{-t}\varphi$ and $\bar\varphi\rightarrow e^{-t}\bar\varphi$ become small and all quantities can be expanded in low powers of the fields.  Thus for $t >>1$,  $\xi \rightarrow 0$ exponentially according to the naive dimension of the power of the fields,
\be
\xi \sim z_1 \varphi e^{-t} + \bar{z}_1 \bar\varphi e^{-t} +z_{11}\varphi\bar\varphi e^{-2t}+\cdots ~~;~~t\rightarrow\infty.
\ee
Hence all renormalization group trajectories flow towards the no radiative correction relation as the theory evolves into the infrared, and they meet at the single trivial fixed point of the theory, $g=1$ and $f=\frac{1}{2}=\bar{f}$ as $t \rightarrow \infty$ .  (Generally we are considering $f\neq 0\neq\bar{f}$.  For $f=0=\bar{f}$, the trivial infrared fixed point is simply $g=1$.  On the other hand if $g=0$, then the action consists only of a free vector superfield, $f=1/2 =\bar{f}$.)

\newsection{The Perturbative Effective K\"ahler Potential}

Similar conclusions can also be drawn from a one loop perturbative analysis of the model given by the action obtained by combining equation (\ref{zeroaction}) with $\kappa =0$ and equation (\ref{KKaction}).  The one loop correction to the dimensionless effective K\"ahler potential is found to be
\bea
\delta K_{\rm 1-loop} &=& -\frac{1}{16\pi^2}\int_0^1 d\xi \ln{\left[\frac{\sinh{\pi R \sqrt{g^2 \xi +r}}}{\sinh{\pi R \sqrt{\xi +r_0}}}\frac{\sinh^2{\pi R \sqrt{\xi}}}
{\sinh^2{\pi R \sqrt{\left( f+\bar{f}\right)\xi}}}\right]} ,
\eea
where $g$, $f$, $\bar{f}$ and $r$ are specified dimensionless tree action terms while the integral is over the four dimensional momentum below the cutoff $\Lambda$, now scaled to $1$.  The fifth dimension K-K modes' propagation has been completely summed over in the loop.

In the four dimensional limit, $ R\rightarrow 0$, with $r =0$, the momentum integral can be explicitly performed to yield the expected result \cite{GRvU}-\cite{PW}
\be
\delta K^{4D}_{\rm 1-loop} = -\frac{1}{16\pi^2}\ln{\left[\frac{g}{f+\bar{f}}\right]},
\ee
which vanishes for the $N=2$ SUSY relationship between the metric and the effective gauge couplings: $g=f+\bar{f}$.  For the case of a noncompact fifth dimension, $ R\rightarrow \infty$, the sum over K-K modes again becomes an integral over the spatial momentum conjugate to the fifth dimension yielding
\bea
 \frac{\delta K_{\rm 1-loop}}{2\pi R} &=& -\frac{1}{32\pi^2} \int_0^1 d\xi \int_{-\infty}^{+\infty} \frac{dk}{2\pi}\ln{ \left[\frac{\left(g^2\xi +k^2\right)\left(\xi +k^2\right)}{\left[ \left(f+\bar{f}\right) \xi +k^2 \right]^2}\right]}\cr
 &=& -\frac{1}{48\pi^2}\left[ g -2\sqrt{f+\bar{f}}+1\right] .
\eea
This vanishes when $g -2\sqrt{f+\bar{f}}+1=0$, which is the identical result as was obtained in the Wilson renormalization group equation analysis. On the other hand, it should be noted that the relation between $g$ and $f+\bar{f}$ for no radiative corrections in this perturbative case differs slightly from that obtained in the Wilson renormalization group analysis for moderate values of $R$. As seen in figure 2, the one loop contributions to the K\"ahler potential increase rapidly with increasing compactification radius $R$.  Indeed, for large radii the quantum corrections are proportional to $R$. Already for $R$ values of order 1 there are substantial deviations from the $R=0$ result. This is readily seen by the shape of the curve on each surface corresponding to the perturbative relation between $g$ and $f+\bar{f}$ for the absence of radiative corrections.  This $\delta K_{\rm 1-loop}=0$ curve is displayed as the darkened solid line in the figure. The grid spacing on 
\begin{figure}
\psfig{file=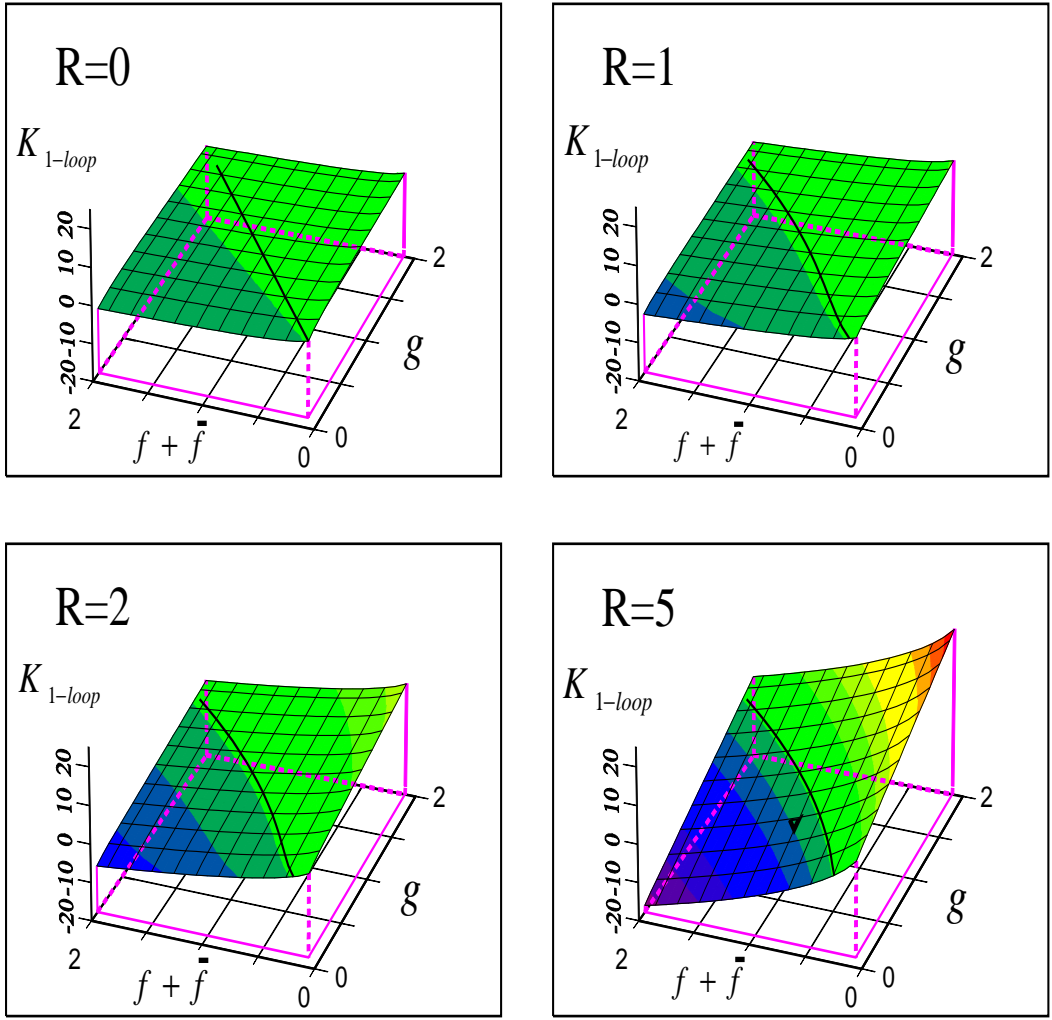,width=6.0in,height=6.0in}
\caption{The one loop radiative corrections to the K\"ahler potential for  $g$ and $f+\bar{f}$ values from 0.1 to 2.0 and for compactification radii $R = 0, 1, 2, 5$. Here $K_{\rm 1-loop}=-16\pi^2 \delta K_{\rm 1-loop}$. The darkened solid line is the no radiative correction relation.}
\end{figure}
the surface corresponds to 0.2 changes in $g$ and $f+\bar{f}$ while the shading of the surface corresponds to changes of 3 in the value of $K_{\rm 1-loop}=-16\pi^2 \delta K_{\rm 1-loop}$.  For $R=0$, the $K_{\rm 1-loop}$ surface is fairly flat, going from a minimum of  -3  to a maximum value of  +3, corresponding to fairly small radiative corrections to $K$.  As $R$ increases to $1$, the $K_{\rm 1-loop}$ values range from -4.79 to 5.67 while for $R=2$, the $K_{\rm 1-loop}$ spread in values is from -7.9 to 10.2.   Thus the range in $K_{\rm 1-loop}$  values increases as $R$ increases.  For $R=5$, the now quite substantial radiative corrections result in the $K_{\rm 1-loop}$ surface ranging from a minimum of -18.3 to a maximum of  +24.8.

The effects of large compact dimensions on the K\"ahler potential can be further demonstrated by considering an explicit tree level potential.  In general, the radiative corrections are small due to the  $\frac{1}{16\pi^2}$ loop factor. As the size of the extra dimension becomes the inverse of this order, however, the loop corrections become comparable with the tree level expression for the K\"ahler potential.  For example, consider the $N=2$ SUSY holomorphic prepotential ${\cal F}$ in four dimensions given by
\be
{\cal F}(\varphi) = \frac{i}{2}(1-2a)\varphi^2 +\frac{ia}{b}(1+b\varphi^2)\ln{(1+b\varphi^2)},
\ee
where $a$ and $b$ are arbitrary parameters. The tree level K\"ahler potential derived from this is simply
\bea
K(\varphi, \bar\varphi) &=& {\rm Im}{\left( {\cal F}_\varphi \bar\varphi\right)}\cr
 &=& \varphi\bar\varphi \left[1 +a\ln{\left( 1+b\varphi^2\right)\left( 1+b\bar\varphi^2\right)}\right]
\label{K}
\eea
while the associated metric is
\be
g(\varphi, \bar\varphi)= 1+a\ln{\left[ \left( 1+b\varphi^2\right)\left( 1+b\bar\varphi^2\right)\right]} +2ab\left[\frac{\varphi^2}{1+b\varphi^2}+\frac{\bar\varphi^2}{1+b\bar\varphi^2}\right] .
\ee
Choosing the effective gauge couplings $f$ and $\bar{f}$ to be related to the metric $g$ according to the $N=2$ SUSY relation $g=f+\bar{f}$, there are no radiative corrections to the K\"ahler potential in four dimensions.  Taking $a=1=b$, the one loop effective K\"ahler potential,
\be
K_{\rm effective} = K +\delta K_{\rm 1-loop} ,
\ee
is plotted in figure 3 for various values of the scalar fields, now taken as real and equal, $\varphi =\bar\varphi$.  
\begin{figure}[h]
\psfig{file=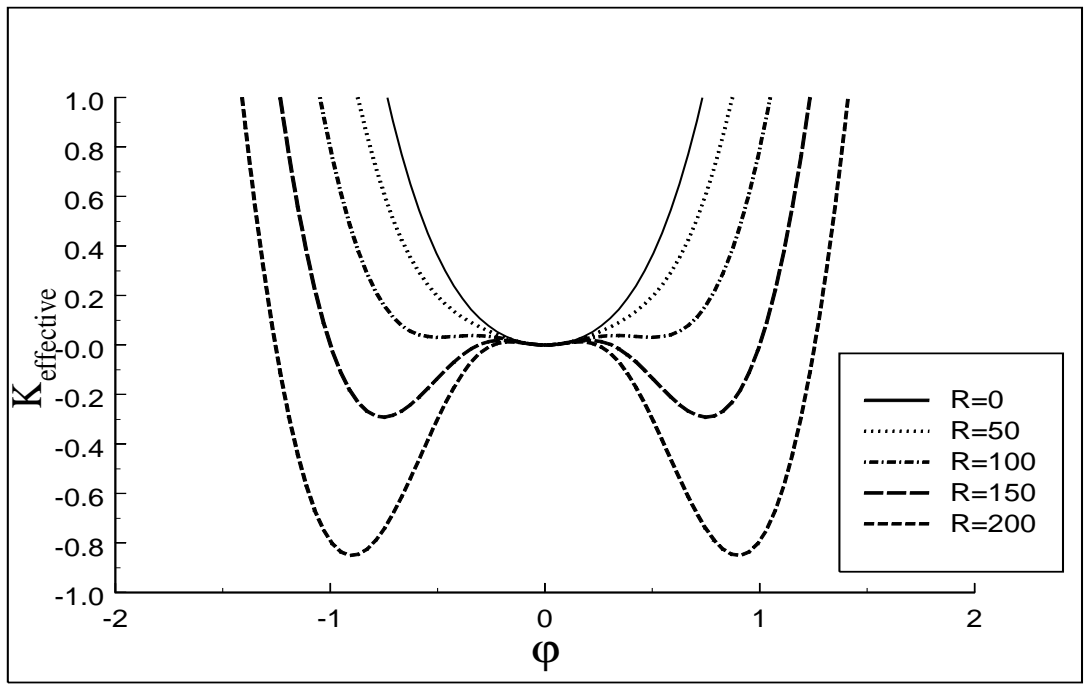,width=6.0in,height=4in}
\caption{$K_{\rm effective}$ as a function of $\varphi $ for compactification radii $R=0, 50, 100, 150, 200$. Note that the $R=0$ curve is simply the tree expression for $K$, equation (\ref{K}).}
\end{figure}
As expected, when $R$ reaches the order of $16\pi^2$, the loop corrections to the K\"ahler potential are qualitatively different than the tree level values even for modest values of the fields. Note that as the scalar field strength continues to increase (beyond that shown in the figure), the effective K\"ahler potential is dominated by its tree value: $K_{\rm effective}\sim \varphi\bar\varphi \ln (\varphi \bar{\varphi})$. This follows since the 1-loop radiative correction to $K$ goes as $g$ and hence as $\ln (\varphi \bar\varphi)$.

\bigskip

This work was supported in part by the U.S. Department 
of Energy under grant DE-FG02-91ER40681 (Task B).
\pagebreak

\newpage
\end{document}